\documentclass[aps,prb,preprint,showpacs,showkeys]{revtex4}
\usepackage{graphicx}
\usepackage{amsfonts}
\usepackage{amsmath}
\usepackage{amssymb}
\usepackage{amsfonts}
\usepackage{graphicx}
\usepackage{bbm}

\begin{document}

\preprint{}
\title[Steady state]{The steady state in noncollinear magnetic multilayers }
\author{Jianwei Zhang}
\affiliation{Department of Physics, 4 Washington Place, New York University, New York,
New York 10003}
\author{Peter M Levy}
\affiliation{Department of Physics, 4 Washington Place, New York University, New York,
New York 10003}

\begin{abstract}
There are at least two different putative steady state solutions for current
across noncollinear magnetic multilayers; one has a discontinuity in the
spin current at the interface the other is continuous. We compare the
resistance of the two and find the solution with the continuous spin
currents is lower.\ By using the entropic principle we can state that this
solution is a better estimate of the resistance for a noncollinear magnetic
multilayer
\end{abstract}

\volumeyear{year}
\volumenumber{number}
\issuenumber{number}
\eid{identifier}
\date{\today}
\published[Published text]{date}
\maketitle






When current is driven across a multilayered structure in which the magnetic
layers are not collinear one has to consider how the spin polarized current
is transmitted between layers. In most treatments one considers the
components of the spin current transverse to the magnetization of a layer is
either absorbed or reflected at the interface upon entering the layer and
that this is the origin of spin torque created by spin currents in magnetic
multilayers \cite{slon-berger}. This scenario does not envisage the
existence of components of the spin accumulation that are transverse to the
magnetization of the layer into which they are injected by the charge
current, but it does assume a steady state of the current is reached, e.g.,
one posits that there is no time dependence of the current densities.\ An
alternate proposal is to inject transverse spin currents through interfaces
that propagate in transition-metal ferromagnets, e.g., in Co up to 3 nm from
the interface, and to have the spatial variation of the spin currents
balanced by transverse spin accumulation thereby achieving steady state.\cite%
{jianwei} In this case the spin currents create an accumulation which acts
as a torque on the background rather than directly suppling angular momentum
to the background magnetization.\cite{spin diffusion} When the current
drives a system out of equilibrium it is not immediately obvious which of
the two scattering processes around the interfaces better describe the
steady state transport across a multilayered structure. To resolve this
question we use the variational principle of transport that says that:
\textquotedblleft in the steady state, the currents are such that the
production of entropy has the largest value consistent with its subsequent
conservation.\textquotedblright\ \cite{ziman} We find in the steady state
transverse spin currents with their attendant accumulations do exist in
magnetic layers in the immediate region about the interface, i.e., within
several nm. Their existence determines the amount of spin torque created on
a magnetic layer for a fixed current; therefore it changes one's predictions
of the critical current needed to switch the magnetization in a multilayered
structure.\cite{expts}

The experiments that have demonstrated current induced magnetization
reversal have been done on nanopillars that consist of two magnetic layers
separated by a nonmagnetic spacer to magnetically isolate the layers, but
otherwise transmit the spin polarized current from one layer to the other,
and nonmagnetic leads.\cite{expts} To resolve the issue of how the spin
polarized current created in one layer is transmitted to another
noncollinear layer we solve for the transport across two magnetic layers and
assume the nonmagnetic spacer is thin enough to be neglected, i.e., when one
drives a\ spin polarized current across the trilayer one uses spacer layer
thicknesses small compared to the spin diffusion length so that one can
coherently transmit spin information from one ferromagnetic layer to the
other;\cite{expts} the analysis can be repeated with a finite spacer
without changing our conclusions. We solve the Boltzmann equation for
current across the layers by using the interface scattering matrix to relate
the distribution functions across their common interface to
self-consistently determine the accumulations and currents in each layer,
e.g., see Waintal and Bauer.\cite{bauer} The additional resistivity due to
the noncollinearity of the layers is found from the change in chemical
potential that develops across the structure, e.g., see Valet and Fert.\cite%
{vf} While this is proportional to the angle between the magnetization of
the layers, it also depends on whether the scattering at the interface is
able to create spin flips; recently we have shown that they are induced by
currents in noncollinear magnetic multilayers.\cite{levy} By determining the
change in chemical potential across the two layered structure we use the
variational principle, that holds that the electrical resistivity is a
minimum in the steady state,\cite{ziman} to discern which scattering at the
interface achieves the lowest resistance and therefore is the best
description for a steady state spin current in noncollinear multilayers.

We consider two bands crossing the Fermi surface of a ferromagnet, a
majority spin band and one minority; this is appropriate for example for Co
where the two additional minority bands could be included but do not add
much to the transport across the interface when the layers are noncollinear.
Our Boltzmann equations for two band are,\cite{jianwei}

\begin{equation}
v_{p}^{x}\cdot\frac{\partial f_{p}}{\partial x}-eE(x)v_{p}\cdot\frac{%
\partial f_{p}^{0}}{\partial\epsilon}=-\frac{1}{\tau_{p}}\cdot\left(
f_{p}-\langle f_{p}\rangle\right) -\frac{1}{\tau_{sf}}\cdot\left(
f_{p}-\langle f_{p^{\prime}}\rangle\right) ,  \label{eq:one}
\end{equation}

\begin{equation}
v_{p}^{x}\frac{\partial f_{p}^{\pm }}{\partial x}\mp iJ\cdot f_{p}^{\pm }=-%
\frac{1}{\tau _{p}^{\prime }}\cdot \left( f_{p}^{\pm }-\langle f_{p}^{\pm
}\rangle \right) ,
\end{equation}%
where $p,p^{\prime }=M$ for majority or $m$ for minority, $J$ is the
exchange splitting between the state on the Fermi surface and its partner
with the same momentum and opposite spin,\cite{jianwei} and $\tau _{p}$\emph{%
\ }and $\tau _{sf}$ represent spin dependent but non-spin-flip and spin-flip
relaxation times. Also, $\tau _{p}^{\prime }$ represent the transverse
relaxation time; in our calculation we take $\tau _{p}^{\prime }=\tau _{p}$
as it does not affect the point we are making in this paper. The
distribution functions $f_{p}(k,x)$ and $f_{p}^{\pm }(k,x)$ are functions of
momentum and position; the angular brackets \ on the right hand side of
Eqs.(1) and (2) denote averages over the Fermi surface. In our notation each
band crossing the Fermi surface is represented by a spinor distribution
function; these are statistical density matrices which characterize the out
of equilibrium state of the current. The first equation refers to the
longitudinal or spin diagonal component on the distribution function on the
Fermi surface, and the current induced \textit{coherences} with states off
the Fermi surface are given by the second equation. We do not write an
equation for the excited state itself as it is not occupied in equilibrium;
this term corresponds to the second longitudinal or diagonal component of
the spinor. As we limit our current analysis to linear response in the
electric field we do not consider it further here.

As we consider two bands, $p=M,m$, there is a spinor for each sheet of the
Fermi surface. For specificity we rewrite the distribution functions as,
\begin{equation}
f_{M}=f_{MM},f_{M}^{+}=f_{MM^{\prime }},f_{M}^{-}=f_{M^{\prime }M},\text{ \
and }f_{m}=f_{mm},\text{\ \ }f_{m}^{+}=f_{m^{\prime
}m},f_{m}^{-}=f_{mm^{\prime }},
\end{equation}%
where the prime denotes an excited state. Note that although there is
scattering on the Fermi surface to produce finite values for $f_{Mm},f_{mM%
}$ we do not consider these components because their contribution to
the current cannot be calculated in the Boltzmann approach as there is no
unique velocity associated with them.\cite{jianwei,stiles-slon} As
we have coherences with excited states of opposite spin, and not only with
states of opposite spin on the Fermi surface, we have in linear response six
components of the distribution functions to keep track of in each layer
which is nearly twice the number when one limits oneself to scattering
\textit{on} the Fermi surface. The boundary condition for these functions at
the interface are,\cite{bauer}
\begin{equation}
f_{qq^{\prime }}^{>}(0^{+})=\sum_{pp^{\prime }}T_{pp^{\prime }qq^{\prime
}}f_{pp^{\prime }}^{>}(0^{-})+\sum_{pp^{\prime }}R_{pp^{\prime }qq^{\prime
}}f_{pp^{\prime }}^{<}(0^{+}),
\label{q02}
\end{equation}

\begin{equation}
f_{qq^{\prime }}^{<}(0^{-})=\sum_{pp^{\prime }}T_{pp^{\prime }qq^{\prime
}}f_{pp^{\prime }}^{<}(0^{+})+\sum_{pp^{\prime }}R_{pp^{\prime }qq^{\prime
}}f_{pp^{\prime }}^{>}(0^{-}),
\label{q01}
\end{equation}%
where $pp^{\prime },qq^{\prime }=MM,MM^{\prime },M^{\prime }M,m^{\prime
}m,mm^{\prime }$ or $mm$. The transmission and reflection coefficients (
probabilities) are
\begin{align}
T_{pp^{\prime }\Rightarrow qq^{\prime }}& =T_{pp^{\prime }qq^{\prime
}}=t_{pq}\ast t_{p^{\prime }q^{\prime }}^{\ast },  \notag \\
R_{pp^{\prime }\Rightarrow qq^{\prime }}& =R_{pp^{\prime }qq^{\prime
}}=r_{pq}\ast r_{p^{\prime }q^{\prime }}^{\ast },  \label{bigtr}
\end{align}%
where $p,p^{\prime },q,q^{\prime }=M,M^{\prime },m^{\prime },m$ , and the
amplitudes $t_{pq},r_{pq}$ should be determined from ab-initio calculations.
\cite{kelly} These coefficients have been determined for interfaces between
ferromagnetic and normal (F/N) metals for states that exist on the Fermi
surface when the system is in equilibrium; in this case states of opposite
spin are coherent in the normal layer but not in the ferromagnet when one
uses a single particle description for the electron states. For this reason
distribution functions in the ferromagnetic layer $f_{qq^{\prime }}$ only
with $q=q^{\prime }$ exist when the components are referred to the axis of
the magnetization of the layer. We have recently shown that to calculate the
conductivity of noncollinear magnetic multilayers by using a local
layer-by-layer approach it is necessary to include additional scattering at
the F/N interfaces; this produces so called current induced coherences in
the ferromagnetic layers, i.e., components of the distribution function $%
f_{qq^{\prime }}$ with $q\neq q^{\prime }$.\cite{levy} In contrast to the
case when one neglects these current induced coherences we have 16
components of $f_{qq^{\prime }}^{\gtrless }$ to consider.

To discuss transport across a noncollinear F/N/F trilayer we write the
transmission and reflection amplitudes and coefficients in terms of the F/N
scattering coefficients discussed above. We will make the simplification of
neglecting the additional scattering at the F/N interfaces arising from
differences in the band structure between the ferromagnetic and normal
layers, i.e., we overlook differences between $k_{n}$ and $k_{M/m}$, but we
do consider $k_{M}\neq k_{m}$; this oversight in no way alters the primary
conclusion we arrive at. When the thickness of the normal spacer is small
compared to the spin diffusion length there is no loss of spin information
and one can write the scattering amplitudes and coefficients connecting the
ferromagnetic layers in terms of those for the F/N interfaces.\cite{bauer}
We use the variational principle cited above to determine whether non
spin-flip or spin-flip scattering at the interface yields the best estimate
for the resistivity. To calculate the angular dependence of the reflection
and transmission due to this perturbation we use the spin rotation matrices
to refer the states quantized along the magnetization of one F layer in
terms of states quantized along the magnetization of the other F layer. In
the limit of neglecting the \textit{non}-spin-flip scattering at the
individual F/N interfaces in which the electron stays on the same sheet of
the Fermi surface we find for the majority state the transmission amplitudes
are

\begin{equation}
t_{MM}=t_{M^{\prime}M^{\prime}}=\cos(\theta/2),
\end{equation}

\begin{equation}
t_{MM^{\prime }}=t_{M^{\prime }M}=iA\sin \theta \sin (\theta /2),
\end{equation}

\begin{equation}
t_{Mm^{\prime}}=t_{M^{\prime}m}=0,
\end{equation}

\begin{equation}
t_{Mm}=t_{M^{\prime }m^{\prime }}=\frac{2k_{M}(1-A\sin \theta )}{%
(k_{M}+k_{m})}(\sin (\theta /2)),
\end{equation}%
where $k_{M},k_{m}$ are the Fermi momenta of the majority and minority
bands, $\theta $ the angle between the magnetic layers, and $A$
parameterizes the amplitude arising from current induced spin-flip
scattering at the F/N interfaces; $A=0$ corresponds to no spin flip
scattering which is the conventional assumption. The half angle terms arise
from the spin rotation matrices while the $\sin \theta $ comes from the
magnitude of the current induced spin-flip scattering.\cite{levy} For the
transmission amplitudes of the minority state we replace $M\rightarrow
m,m\rightarrow M,$ $k_{M}\rightarrow k_{m}$, and $A\rightarrow B$ . When we
include the scattering at the F/N interfaces that comes from $k_{M/m}\neq
k_{n},$ $t_{MM}$ and $t_{mm}$ are less than $1$ when $\theta =0$, but this
does not change the conclusions we arrive at here.\cite{scattering}

For the reflection amplitudes we write for the majority state
\begin{equation}
r_{MM}=r_{M^{\prime }M^{\prime }}=\frac{k_{M}-k_{m}}{k_{M}+k_{m}}(1-A\sin
\theta )(1-\cos \theta ),
\end{equation}

\begin{equation}
r_{MM^{\prime }}=r_{M^{\prime }M}=-i\frac{k_{M}-k_{m}}{k_{M}+k_{m}}A\sin
^{2}\theta ,
\end{equation}

\begin{equation}
r_{Mm^{\prime }}=r_{Mm}=r_{M^{\prime }m^{\prime }}=r_{M^{\prime }m}=0,
\end{equation}%
while those for the minority state are obtained by the following
replacements $M\rightarrow m,m\rightarrow M,$ $k_{M}\rightarrow k_{m}$, and $%
A\rightarrow B$ .The unitary of scattering matrix can be confirmed as $%
\sum_{qq^{\prime }}T_{pp^{\prime }qq^{\prime }}+\sum_{qq^{\prime
}}R_{pp^{\prime }qq^{\prime }}=1$. Based on a specific mechanism, e.g., the
current induced spin-flip suggested in Ref.(8), the values for $A$ and $B$
can be determined, but their precise values are unimportant to our argument;
\textit{only their existence is crucial}.

\begin{figure}[tbp]
\centering \includegraphics[height=12cm]{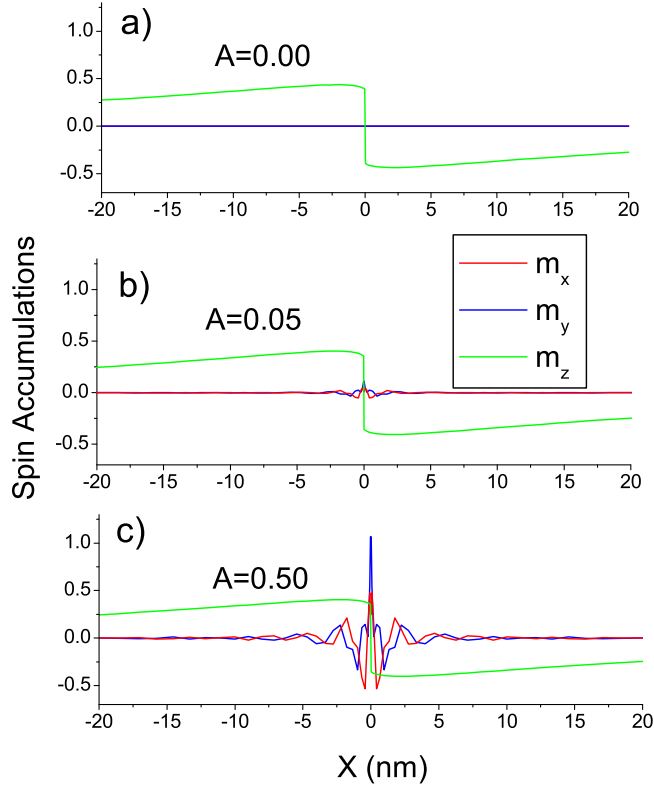}
\caption{The spin accumulations for two Co layers for $\protect\theta =90^{0}
$ calculated by solving Eqs.(1) and (2) and using the transmission and
reflection amplitudes and coefficents given in the text. In (a) we consider
no spin-flip scattering the interface, $A=B=0$; while there is longitudinal
spin accumulation $m_{z}$ (parallel to the \textit{local} magnetization) the
transverse (to the \textit{local} magnetization) is zero, i.e., there is no
injection of transverse spin currents in the absence of the interfacial
spin-flip scattering. In (b) we have a very small amount of spin-flip
scattering $A=B=0.05$ and find a small amount of transverse spin
accumulation $m_{x}$ and $m_{y}$ while the longitudinal is practically
unchanged. For spin-flip scattering ten times stronger $A=B=0.5$ the
transverse accumulation is proportionately larger, but the range is
unchanged (see Fig.2),; the longitudinal remains unchanged.}
\label{fig01}
\end{figure}

\begin{figure}[tbp]
\centering \includegraphics[height=7cm]{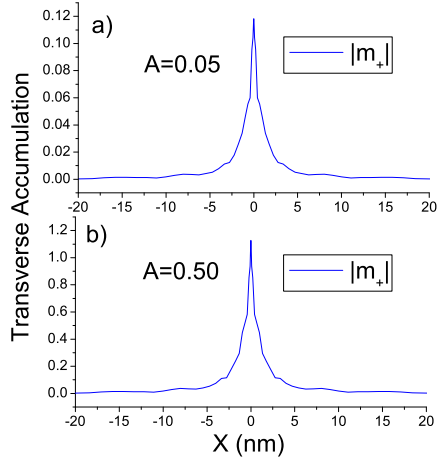}
\caption{The modulus of the transverse accumulation $m_{+}\equiv m_{x}+im_{y}
$ for the two values of the spin-flip scattering considered in Fig.1. Note
the scale change for the accumulation by a factor of ten; otherwise the two
accumulations are very similar. This underscores that the length scale of
the transverse spin accumulation is determined by the exchange parameter $J$
in the equation of motion for the transverse distribution function (see
Eq.(2)), and not by the amount of spin-flip scattering at the interfaces.}
\label{fig02}
\end{figure}

\begin{figure}[tbp]
\centering \includegraphics[height=12cm]{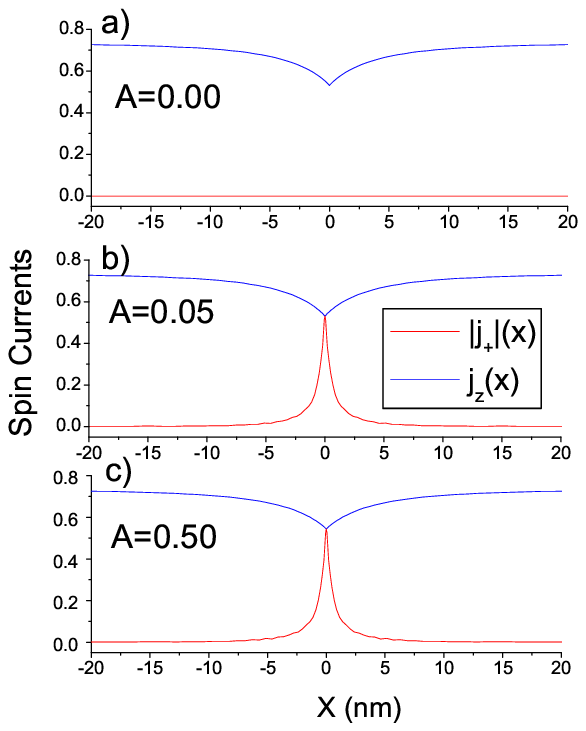}
\caption{The spin currents across two noncollinear Co layers with $\protect%
\theta =90^{0}$ calculated from the solutions for the distribution functions(see
Eqs.(1) and (2)); $j_{z}$is the component parallel to the \textit{local}
magnetization, and $j_{+}\equiv j_{x}+ij_{y}$ is transverse. The
longitudinal and transverse spin currents are unchanged as we vary the
amount of spin-flip scattering at the interface, albeit for $A=B=0$ there is
no injection of transverse spin current.}
\label{fig03}
\end{figure}

To determine whether current induced spin-flip scattering is needed to
achieve steady state current conditions we compared the resistance for $A=B=0
$ to the resistance one finds for $A=B\neq 0$ ; as our result is not
sensitive to the values for them, in our calculations we consider two set of
values $A=B=0.05$ and $0.5$. Also we used the Fermi momenta $k_{M},k_{m}$
for Co \cite{zahn} so that $\frac{k_{M}-k_{m}}{k_{M}+k_{m}}\simeq 0.4$. We
solved Eqs.(1) and (2) for the out of equilibrium spin transport across two
noncollinear magnetic Co layers by using the above transmission and
reflection amplitudes and the coefficients Eqs.(\ref{bigtr}) to match the
out of equilibrium distribution functions across the F/N/F interface. In
Figs. 1 through 3 we compare the spin accumulations and currents across the
interfacial region for the three sets of $A=B$ , for $\theta =90^{0}$. For $%
A=B=0$ there is no transverse accumulation nor current ($j_{+}\equiv
j_{x}+ij_{y})$; only the longitudinal component of the spin current $j_{z}$
exists in the ferromagnetic layers as expected. For $A=B\neq 0$ the
transverse spin current exists and is independent of the amount of spin-flip
scattering at the interface; only the amount of transverse accumulation
depends on the $A$ and $B$. From Fig.2 we see that the length scale of the
transverse accumulation ($m_{+}\equiv m_{x}+im_{y}$) does not depend on $A=B$%
; rather it is controlled by the exchange splitting $J$ in Eq.(2) for the
transverse distribution function. Note that the\ vertical scales for the
accumulation in Figs.2 differ by a factor of ten, so that the transverse
accumulation is nearly linear dependent on the spin-flip scattering
parameter $A$.

The additional resistance due to noncollinearity of layers is found from the
additional voltage drop $\Delta V$ of the electrochemical potential; see
Valet and Fert.\cite{vf} In the present notation this electrochemical
potential $\mu _{e}$ is defined as $\left\langle f_{MM}\right\rangle
+\left\langle f_{mm}\right\rangle $ where the brackets denote averages over
the Fermi surface, and the additional resistance is
\begin{equation}
\Delta R=\frac{\Delta V}{j_{e}}=\frac{1}{j_{e}}\int_{-\infty }^{+\infty
}\left( \frac{1}{e}\frac{\partial \mu _{e}}{\partial x}-E_{0}\right) dx,
\label{resis}
\end{equation}%
where $j_{e}=\frac{4\pi m^{2}}{3\hbar ^{2}}e^{2}E_{0}\lambda \left(
v_{M}^{2}+v_{m}^{2}\right) $, $\lambda $\ is the mean free path in the bulk
of the metal, and for Co the ratio of the Fermi velocities is $%
v_{M}:v_{m}=1:0.78$ . While the integral is over the entire structure we can
cut it off when the chemical potential is independent of position. As the
spin diffusion length for a typical 3d transition-metal, e.g., Co, is of the
order of $60$ nm we have limited the integration to $90$ nm about the
interface. We have calculated the resistances for the three sets of $A$ and $%
B$ as a function of angle and present our results in Fig.4. While the
lowering of the resistance is proportional to the amount of current induced
spin-flip scattering ($A$ and $B$) the signal conclusion is that as long as $%
A$ and $B$ are nonzero the resistance is always lower than for $A=B=0.$ We
conclude that when the spin current is discontinuous at the F/N interfaces,
i.e., $A=B=0$ the resistance is always larger than when the current is
continuous ($A=B\neq 0$). By invoking the principle that the electrical
resistance is a minimum in the steady state of the current, we conclude that
the spin-flip interface scattering is necessary to achieve steady state
conditions.

\begin{figure}[tbp]
\centering \includegraphics[height=9cm]{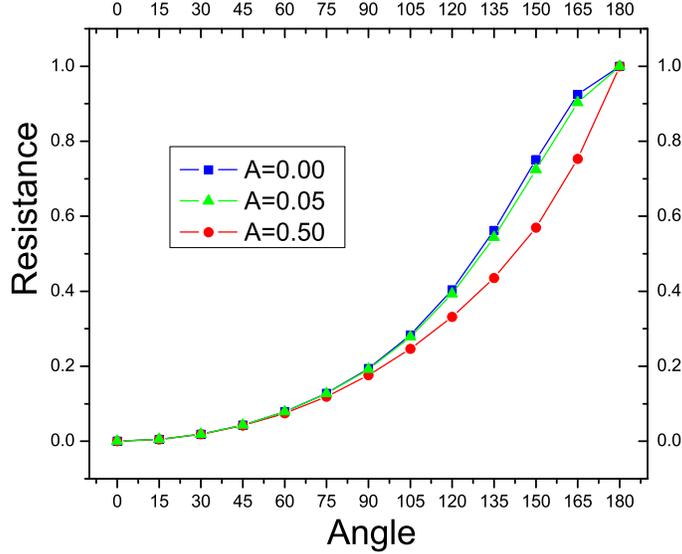}
\caption{The resistance across two Co layers, each 90 nm, as determined from
Eq.(\ref{resis}) as a function of the angle between their magnetizations. As we
neglect the \textit{non}-spin-flip scattering at the F/N interfaces the resistance at
$\protect\theta =0$ is zero and that for $\protect\theta =180^{0}$ is
independent of the spin-flip scattering $A=B$ at the interface; all
resistances are normalized to that at $\protect\theta =180^{0}$. The salient
feature is that the presence of spin-flip scattering at the interface
invariably lowers the resistance.}
\label{fig04}
\end{figure}

The resistance shown in Fig.\ref{fig04} can be written as

\begin{equation}
R=R_{0}+R_{\theta }\ast f(\theta ),
\end{equation}%
where $R_{0}$ is the resistance coming from the band mismatch at the F/N
interfaces, i.e., $k_{M/m}\neq k_{n}$; we have set this to zero in our
calculation. For $A=B=0$ we are able to fit our results to the form
suggested by\cite{pratt}

\begin{equation}
f(\theta )=\frac{1-\cos ^{2}\theta /2}{1+\chi \cos ^{2}\theta /2},
\end{equation}%
and we find $\chi =3.4$. For $A=B=0.05,$ $\chi =3.7$; while for $A=B=0.5$
this form does not provide a suitable fit to the angular dependence of the
resistance. However when we use

\begin{equation}
f(\theta )=\frac{1-\cos \theta /2}{1+\chi \cos \theta /2},
\end{equation}%
we are able to obtain an excellent fit with $\chi =1.07$ for $A=B=0.5$.

In equilibrium only states up to the Fermi level are occupied and the
scattering at the F/N interfaces is characterized by $A=B=0$. When the
current is switched on the current induced spin-flip scattering at the
interfaces induces coherences between states on the Fermi surface and
excited states of opposite spin. This injection of transverse components of
the spin current is offset by the interference coming from the individual $k$
states on the Fermi surface,\cite{fert-stiles} so that after a
characteristic length $\lambda _{tr}$ the current is parallel to the
background magnetization; from Fig.3 we can see that $\lambda _{tr}\simeq 3nm
$.\cite{spin diffusion} While in linear response we are only interested in
the coherences $f_{pp^{\prime }}$ between the spin states on the Fermi
surface $p$ and the excited states of opposite spin $p^{\prime }$, it
follows from the inequality for density matrices representing statistical
mixtures of states that

\begin{equation}
f_{pp}f_{p^{\prime }p^{\prime }}\geq \left\vert f_{pp^{\prime }}\right\vert
^{2},  \label{??}
\end{equation}%
i.e., the presence of coherences indicate that there is an occupancy of the
excited state when the system is out of equilibrium; albeit it is a \textit{%
higher order effect}. As the transverse components of the spin current\
disappear close to the interface (within 3-5 nm\cite{spin diffusion}) the
time required to achieve this \textquotedblleft
equilibrium\textquotedblright\ is short compared to the spin-flip time that
equilibrates the longitudinal components of the spin current.

The existence of finite current induced coherences $f_{pp^{\prime }}$ close
to the interface give rise to the transverse spin accumulations found in
this and previous studies; it should be stressed that our self consistent
solutions of the Boltzmann equation do not mandate transverse spin
accumulation, rather they are found to be necessary concomitant to achieve a
steady state in noncollinear magnetic structures.\cite{spin diffusion} It
is in this sense that our calculations indicate that spin-flip scattering at
the interfaces is necessary to achieve steady state current conditions in a
noncollinear multilayered structure. Finally, the time necessary for
creating the transverse spin accumulation is short compared to that for the
longitudinal; this is controlled by the spin-flip time which for Co is of
the order of $10^{-12}$ sec. Both times are short compared to the time it
takes for the background magnetization to move, $\sim 10^{-9}$ sec.;
therefore one can achieve steady state currents in noncollinear magnetic
structures prior to magnetization reversal \textit{provided one allows for
transverse spin accumulation} in the regions about the interfaces.
Conversely, if one does not allow this one cannot achieve steady state
conditions in noncollinear magnetic structures.

In conclusion, in the calculations whose results we presented we have taken
into account differences in the band structures in the ferromagnetic layers,
i.e., $k_{M}\neq k_{m}$. In the conventional approach, $A=B=0$, there is a
discontinuity in the spin current at the N/F interfaces, while in our
approach, $A=B\neq 0$, there is none. Therefore the origin of the
discontinuity cannot be the band mismatch\textit{ per se}; rather the key
difference between the two scenarios is that we account for the coherence
between states of opposite spin in the ferromagnetic layers that is lost in
effective single electron treatments. When we account for these coherences
in our model with spin split bands in the ferromagnetic layers the spin
currents are close to those found in previous calculations based on unsplit
free electron bands (which maintain coherences throughout the multilayer).

We would like to acknowledge very helpful discussions with Jerry Percus on
the use of the resistivity minimum principle for steady state currents, and
to Shufeng Zhang on general transport theory. This work done under a grant
from the National Science Foundation(Grant DMR 0131883).


\begin{thebibliography}{99}
\bibitem{scattering} For the F/N/F trilayer we can take into account the
non-spin-flip scattering at the F/N interfaces, i.e., $k_{M/m}\neq k_{n}$.
If we neglect multiple reflections.between the interfaces the transmission
amplitudes connecting the two ferromagnetic layers can be written as

\begin{equation}
t_{MM}=\frac{4k_{M}k_{n}}{(k_{M}+k_{n})^2}\cos(\theta/2),
\end{equation}

\begin{equation}
t_{MM^{\prime }}=i\frac{2k_{M}}{(k_{M}+k_{n})}A\sin \theta \sin (\theta /2),
\end{equation}

\begin{equation}
t_{Mm^{\prime }}=0,
\end{equation}%
and

\begin{equation}
t_{Mm}=\frac{4k_{M}k_{n}(1-A \sin \theta)}{(k_{M}+k_{n})(k_{m}+k_{n})}\cos
(\theta /2).
\end{equation}%
From these we can see that the trilayer scattering has the same angle
dependence as indicated above for $k_{M/m}=k_{n}$ . When we include the
angular dependence coming from multiple reflections they enter in the
demominator of the transmission amplitudes ( see Ref.6), and complicate its'
angular dependence. However for the 3d transition-metal normal metal
interfaces we consider, e.g., Co/Cu, the dominant transmission is carried by
the majority band and the reflection in this channel is small; therefore the
correction due to multiple reflections is quite small.
\end{thebibliography}
\end{document}